\documentclass[pre,aps,preprint,amsmath,superscriptaddress]{revtex4-1}
\usepackage{natbib}
\usepackage{graphicx}
\usepackage{dcolumn}
\usepackage{txfonts}
\usepackage{bm}

\begin{document}

\title{Convective Heat Transfer in Porous Materials}
\author{Peng Jin}\email{19110190022@fudan.edu.cn}
\affiliation{Department of Physics, State Key Laboratory of Surface Physics, and Key Laboratory of Micro and Nano Photonic Structures (MOE), Fudan University, Shanghai 200438, China}

\author{Gaole Dai}
\affiliation{School of Sciences, Nantong University, Nantong 226019, China}

\author{Fubao Yang}
\affiliation{Department of Physics, State Key Laboratory of Surface Physics, and Key Laboratory of Micro and Nano Photonic Structures (MOE), Fudan University, Shanghai 200438, China}

\date{\today}

\begin{abstract}

Thermal convection stands out as an exceptionally efficient thermal transport mechanism, distinctly separate from conduction and radiation. Yet, the inherently elusive nature of fluid motion poses challenges in accurately controlling convective heat flow. While recent innovations have harnessed thermal convection to achieve effective thermal conductivity, fusing thermal convection in liquids and thermal conduction in solids together to form hybrid thermal metamaterials is still challenging. In this review, we introduce the latest progress in convective heat transfer. Leveraging the right porous materials as a medium allows for a harmonious balance and synergy between convection and conduction, establishing stable heat and fluid flows. This paves the way for the innovative advancements in transformation thermotics. These findings demonstrate the remarkable tunability of convective heat transport in complex multicomponent thermal metamaterials.


\end{abstract}

\maketitle

\section{Introduction}

Over the past decade, the emergence of thermal metamaterials~\citep{PJ-DaiIJHMT20,PJ-rmp,PJ-click,PJ-research,PJ-am,PJ-zhounc,PJ-XuSCPMA20,PJ-Huang20,PJ-WangPRE20,PJ-XuPRAP20,PJ-XuESEE20,PJ-HuangESEE20,PJ-XuEPJB20,PJ-SuEPL20,PJ-WangPRAP20,PJ-XuIJHMT20,PJ-XuAPL20,PJ-XuEPL20,PJ-XuCPLEL20,PJ-WangICHMT20,PJ-HuangPhysics20,PJ-YangJAP20,PJ-JinIJHMT20,PJ-XuEPL20-1,PJ-XuPRE20,PJ-WangiScience20,PJ-YangPRAP20,PJ-XuCPL20,PJ-XuIJHMT21,PJ-DaiJNU21,PJ-DaiiScience21,PJ-WangEPL21,PJ-XuEPL21,PJ-LiuJAP21,PJ-WangATE21,PJ-WangPRAP21,PJ-XuPRE21,PJ-YangPR21,PJ-JinIJHMT21,PJ-ZhangTSEP21,PJ-TianIJHMT21,PJ-XuAPL21} and transformation thermotics~\citep{PJ-FanAPL2008,PJ-XuEPL21-1,PJ-LiuJAP21-1,PJ-LeiEPL21,PJ-QuEPL21,PJ-GaoNM21,PJ-ZhangATS22,PJ-HuangAMT22,PJ-ZhuangSCPMA22,PJ-DaiPRAP22,PJ-XuPRL22,PJ-LinSCPMA22,PJ-ZhangCPL22,PJ-WangCPB22,PJ-YangPRAP22,PJ-LiPF22,PJ-ZhuangPRE22,PJ-JinPNAS23,PJ-XuBook23,PJ-XuPRL22-1,PJ-YaoISci22,PJ-ZhangPRD22,PJ-ZhouEPL23,PJ-ZhangPRA23,PJ-LeiIJHMT23} has greatly broadened the horizons of heat manipulation~\citep{PJ-ZhuangIJMSD23,PJ-XuNSR23,PJ-LeiMTP23,PJ-ZhangCPL23,PJ-YangPRA23,PJ-ZhangNRP23,PJ-XuPANS23,PJ-DaiPR23,PJ-LiJAP10,PJ-QiuEPL13,PJ-QiuIJHT14,PJ-QiuAIPAdv.15,PJ-Tan20,PJ-Tan16,PJ-Tan15,PJ-Tan11,PJ-Tan4,PJ-ShenAPL16,PJ-ShenPRL16,PJ-XinPA17,PJ-HuangFP17,PJ-HuangPB17,PJ-XinFP17,PJ-YangAPL17,PJ-XuEPJB17}. This expansion has proven invaluable in a variety of applications~\citep{PJ-WangJAP17,PJ-JiPA18,PJ-MengCPB18,PJ-ShangIJHMT18,PJ-JiCTP18,PJ-DaiEPJB18,PJ-DaiPRE18,PJ-WangJAP18,PJ-ShangJHT18,PJ-WangIJTS18,PJ-JiIJMPB18,PJ-XuEPJB18,PJ-XuJAP18,PJ-XuPLA18,PJ-XuPRE18,PJ-DaiJAP18,PJ-HuangPP18,PJ-ShangAPL18,PJ-YangJAP19,PJ-XuEPJB19,PJ-XuPRE19,PJ-WangPRA19,PJ-XuPRA19a,PJ-XuEPL19,PJ-YangPRE19,PJ-XuPRA19,PJ-YangEPL19,PJ-ZhouESEE19,PJ-XuPRAP19,PJ-XuEPJB19-1,PJ-YangEPL19-1,PJ-XuPRE19-1,PJ-HuangESEE19}, from thermal cloaking and camouflage~\citep{PJ-XuSCPMA20,PJ-Huang20,PJ-WangPRE20,PJ-XuPRAP20,PJ-XuESEE20,PJ-HuangESEE20,PJ-XuEPJB20,PJ-SuEPL20,PJ-WangPRAP20,PJ-XuIJHMT20,PJ-XuAPL20,PJ-XuEPL20,PJ-XuCPLEL20,PJ-WangICHMT20,PJ-HuangPhysics20,PJ-YangJAP20,PJ-JinIJHMT20,PJ-XuEPL20-1,PJ-XuPRE20,PJ-WangiScience20,PJ-YangPRAP20,PJ-XuCPL20,PJ-XuIJHMT21,PJ-DaiJNU21,PJ-WangEPL21,PJ-XuEPL21,PJ-LiuJAP21,PJ-WangATE21,PJ-WangPRAP21,PJ-XuPRE21,PJ-ZhuangIJMSD23,PJ-XuNSR23,PJ-LeiMTP23,PJ-ZhangCPL23,PJ-YangPRA23,PJ-ZhangNRP23,PJ-YangPRE19,PJ-XuPRA19} to heat management in microchips~\cite{PJ-YaoISci22,PJ-ZhouEPL23,PJ-LeiIJHMT23}, energy conservation in everyday life~\citep{PJ-JiCTP18,PJ-DaiEPJB18,PJ-DaiPRE18,PJ-WangJAP18,PJ-ShangJHT18,PJ-WangIJTS18}, and thermoregulation in biological cells~\cite{PJ-ZhangPRA23,PJ-ZhangNRP23}. Yet, much of the progress in this field has been concentrated on conductive thermal metamaterials~\citep{PJ-FanAPL2008,PJ-LiJAP10,PJ-QiuEPL13,PJ-QiuIJHT14,PJ-QiuAIPAdv.15,PJ-Tan16,PJ-ShenPRL16}. These materials primarily rely on diffusive or effective heat conduction, which is constrained by Onsager's reciprocity. Such reliance places constraints on the versatility of heat manipulation. Additionally, traditional thermal metamaterials lack the flexibility to adjust their functions based on given temperature conditions~\citep{PJ-Tan16,PJ-ShenPRL16}, depriving them of the adaptive control often required.

Thermal convection~\citep{PJ-DaiPRE18,PJ-DaiPRE23,PJ-DaiJAP18,PJ-YangESEE19,PJ-JinPNAS23,PJ-XuSCPMA20,PJ-XuIJHMT20}, with its distinct nature, plays a crucial role as a mechanism for thermal transport. Historically, its role was often overshadowed in the realms of thermal metamaterials and transformation thermotics. Only recently has the theory of transformation thermotics expanded its scope to include thermal convection~\citep{PJ-DaiPRE18,PJ-DaiPRE23,PJ-DaiJAP18,PJ-XuIJHMT20}, necessitating the creation of a novel theoretical framework. Yet, merging thermal convection in liquids with thermal conduction in solids to create hybrid thermal metamaterials presents a significant challenge. This is because these distinct paths of heat transport must align and collaborate harmoniously to generate stable heat and liquid flows that fulfill the requirements of the underlying thermotic transformation. Designing such hybrid materials proves more intricate than traditional all-solid thermal metamaterials. While there have been efforts to integrate thermal convection to achieve remarkable thermal conductivity levels (exemplifying synthetic Onsager reciprocity)~\cite{PJ-31,PJ-32,PJ-33}, there is a pressing need to develop thermal metamaterials that can simultaneously control both conductive and convective heat flows beyond the bounds of Onsager reciprocity.

In this review, we present the latest advancements in heat transfer of porous materials. We commence by elaborating on the foundational principles of transformation thermotics, addressing both steady-state and transient-state challenges of convective heat transfer in porous mediums~\cite{PJ-DaiPRE18,PJ-DaiPRE23,PJ-DaiJAP18}. This potent theory paves the way for the conceptual design of innovative thermal devices, including illusion and camouflage mechanisms~\cite{PJ-YangESEE19,PJ-XuSCPMA20}. Further, we clarify the emergence of experimental platforms for the realization of continuous switch between thermal cloak and thermal concentration~\cite{PJ-JinPNAS23}, which reveals the significant tunability of the hybrid thermal metamaterial. Finally, we envision that such porous mediums could serve as the ideal physical platform for achieving robust thermal-protected transport with topological features.

\section{Steady-state Transformation Thermo-hydrodynamics}

	When addressing heat transfer in fluids, we begin by adjusting the heat conduction equation for incompressible flow, excluding heat sources and disregarding the viscous dissipation term~\cite{PJ-Landau}, as
	\begin{equation}\label{PJ-heat transfer}
	\rho C_{p}\nabla\cdot (\vec{v}T)=\nabla \cdot(\eta\nabla T),
	\end{equation}
	where $\rho$, $C_{p}$, $\eta$, and $\vec{v}$ are respectively the density, specific heat at constant pressure, thermal conductivity, and the velocity of the fluid. As is known, $ \rho C_{p}\nabla\cdot (\vec{v}T) $ is the term due to advection. Equation (\ref{PJ-heat transfer}) represents the convection-diffusion equation. For the sake of clarity, we assume a laminar, Newtonian flow and consider the density to be unaffected by temperature variations.
	
	
	
	For the coordinate transformation $\{x_{i}\}\rightarrow\{y_{j}\}$ and the associated Jacobian matrix $\textbf{J}=\frac{\partial(y_{1},y_{2},y_{3})}{\partial(x_{1},x_{2},x_{3})},$ we can write~\cite{PJ-JRCI2013}
	\begin{equation}
	\rho C_{p} \sum\limits_{j}\frac{\partial }{\partial y_{i}} \left (\frac{1}{\det\textbf{J}}\sum\limits_{i}J_{ij}^{\intercal}v_{i}T\right ) =\sum\limits_{ijkl}\frac{\partial}{\partial y_{k}}\left (\frac{1}{\det\textbf{J}}J_{ki}\eta_{ij}J_{jl}^{\intercal}\frac{\partial T}{\partial y_{l}}\right ) .
	\end{equation}
	Let $ \vec{v}'=\frac{\textbf{J}\vec{v}}{\det\textbf{J}} $
	and $ \eta'=\frac{\textbf{J}{\eta}\textbf{J}^{\intercal}}{\det\textbf{J}}, $ and we achieve
	\begin{equation}\label{PJ-heat transfer transformed}
	\rho C_{p}\left [\nabla'\cdot(\vec{v}'T)\right ]=\nabla'\cdot(\eta'\nabla'T).
	\end{equation}
	 Eqs.~$\left(\ref{PJ-heat transfer}\right)$~and~$\left(\ref{PJ-heat transfer transformed}\right)$ have the consistent form, and thermal convection are included in transformation thermotics.
	
	Following this, we conceptualize and determine the velocity distribution $\vec{v}'(\vec{r},t)$ and the anisotropic thermal conductivity $ \eta' $ of the liquid medium. Typically, to fully characterize the state of fluids, we require knowledge of the velocity $ \vec{v} $ and two additional thermodynamic quantities, such as $ \rho $ and pressure $ p $. These parameters are ascertained using Eq.~$\left(\ref{PJ-heat transfer}\right)$ in conjunction with the Navier-Stokes equations and the continuity equation~\cite{PJ-Landau}
	\begin{equation}\label{navier stokes}
	(\vec{v}\cdot\nabla)\vec{v}=-\frac{1}{\rho}\nabla p+\frac{\beta}{\rho}\nabla\cdot\nabla\vec{v},
	\end{equation}	
	\begin{equation}\label{PJ-conservation law}
	\nabla\cdot\vec{v}=0.
	\end{equation}
	Here, $ \beta $ denotes the dynamic viscosity. For clarity, we take $ \vec{v}(\vec{r},t)=\vec{v}(\vec{r}) $ and $ \rho(\vec{r},t)\equiv\rho $. However, Eq.~$\left(\ref{PJ-conservation law}\right)$ retains its form under coordinate transformation whereas Eq.~$\left(\ref{navier stokes}\right)$ typically does not. But we can overlook nonlinear term $ (\vec{v}\cdot\nabla)\vec{v} $ when Reynolds number Re is small (akin to the elastic equation in~\cite{PJ-NJP2006}). Experimentally, inducing anisotropy in $ \eta' $ for fluids poses challenges, even though this has been effectively achieved for heat conduction in solids. Encouragingly, recent progress in velocity control, as highlighted in \cite{PJ-PRL2011}, spurs us to simultaneously explore heat transfer and velocity management in porous media.
	
	In fully-filled porous media, we give equations for steady flow as~\cite{PJ-Bear1,PJ-Bear2}
	\begin{equation}\label{PJ-heat transfer porous}
	\rho_{f} C_{p,f}(\vec{v}\cdot \nabla T)=\nabla \cdot(\eta_{m} \nabla T),
	\end{equation}
	\begin{equation}\label{PJ-darcy's law}
	\nabla p+\frac{\beta}{k}\vec{v}=0,
	\end{equation}
	\begin{equation}\label{PJ-conservation law 2}
	\nabla\cdot\vec{v}=0,
	\end{equation}
	where $ k $ denotes the permeability and $ \eta_{m} $ is the effective thermal conductivity of the porous media. Meanwhile, $ \rho_{f} $ and $ C_{p,f} $ are the density and specific heat at constant pressure of fluid material, respectively. By taking the volume average of solid and liquid components~\cite{PJ-Bear2}, the effective conductivity $ \eta_{m} $ is given by
	\begin{equation}
	\eta_{m}=(1-\phi)\eta_{s}+\phi\eta_{f}.
	\end{equation}
	where $ \phi $ represents the porosity and $\eta_{f} $ and $\eta_{s} $ are the thermal conductivity of fluid and solid material of porous media, respectively. In Eq.~$\left(\ref{PJ-heat transfer porous}\right),$ the local thermal equilibrium of fluids and solid materials is assumed, indicating that they possess the same temperature at the contact point. Meanwhile, we assume $ \nabla\cdot (\vec{v}T)=\vec{v}\cdot\nabla T $, given by Eq.~$\left(\ref{PJ-conservation law 2}\right).$ Eq.~$\left(\ref{PJ-darcy's law}\right)$ denotes the Darcy's law, in case of small-enough Re and $ k $. From $\lambda=-\frac{k}{\beta}$ and $\vec{v}'=\textbf{J}\vec{v}/(\det\textbf{J}) $, we easily rewrite Eq.~$\left(\ref{PJ-darcy's law}\right)$ under transformation $\{x_{i}\}\rightarrow\{y_{j}\}$,
	\begin{equation}
	v'_{j}=\sum\limits_{i}J_{ji}v_{i}/(\det\textbf{J})=\sum\limits_{ik}J_{ji}\lambda_{ki}\frac{\partial p}{\partial x_{k}}/(\det\textbf{J})=\sum\limits_{ikl}J_{lk}\lambda_{ki}J^{\intercal}_{ij}\frac{\partial p}{\partial y_{l}}/(\det\textbf{J}),
	\end{equation}
	which indicates the relationships: $ \vec{v}'=\lambda'\nabla'p $ and $ \lambda'=\frac{\textbf{J}\lambda\textbf{J}^{\intercal}}{\det\textbf{J}}. $
	
	All the Eqs.~$\left(\ref{PJ-heat transfer porous}\right)$, $\left(\ref{PJ-darcy's law}\right)$ and $\left(\ref{PJ-conservation law 2}\right)$ keep invariant form under any coordinate transformation, and we can obtain the wanted temperature and velocity distribution without tuning properties of fluid materials. That is to say, we only need to transform the permeability
	\begin{equation}
	k'=\frac{\textbf{J}k\textbf{J}^{\intercal}}{\det\textbf{J}},
	\end{equation}
	and the thermal conductivity
	\begin{equation}
	\left\{\begin{array}{ll}\eta'_{m}=\dfrac{\textbf{J}\eta_{m}\textbf{J}^{\intercal}}{\det\textbf{J}},\\
	\eta_{f}'=\eta_{f},\\
	\eta_{s}'=\dfrac{\eta_{m}'-\phi\eta_{f}}{1-\phi}.
	\end{array}\right.
	\end{equation}
	Using different spatial transformations, we can manipulate the heat flow as we wish.

\section{Transient-state Transformation Thermo-hydrodynamics}

In this section, we extend the transformation theory to encompass transient thermal convection in porous media. The Darcy's law can be generalized as follows~\cite{PJ-BurcharthCE95,PJ-ZhuTPM2014,PJ-ZhuTPM2016,PJ-DaiJAP18}:
\begin{equation}\label{PJ-e0}
\tau\frac{\partial \vec{v}}{\partial t}+\vec{v}=-\dfrac{\beta}{\eta}\nabla p.
\end{equation}
Here, $ \tau $ denotes the characteristic time measuring the velocity varying. Additionally, $ \beta $ signifies the permeability of porous medium, while $ \eta $ stands for the dynamic viscosity. In many scenarios, the relaxation process within porous media is rapid, resulting in the term $ \tau\frac{\partial \vec{v}}{\partial t} $ being quite negligible~\cite{PJ-ZhuTPM2016}. Consequently, we can omit this term and continue to operate within the steady Darcy framework. Furthermore, the continuity equation can be modified as~\cite{PJ-Landau}
\begin{equation}\label{PJ-e1}
\frac{\partial (\phi\rho_{f})}{\partial t}+\nabla\cdot(\rho_{f}\vec{v})=0,
\end{equation}
where $ \rho_{f} $ denotes the density of the fluid medium and $ \phi $ stands for the porosity. Then, the heat transfer of incompressible flow in fully-filled porous media is given by~\cite{PJ-Bear1,PJ-Bear2}
\begin{equation}\label{PJ-e2}
\frac{\partial (\rho C)_{m} T}{\partial t}+\nabla\cdot(\rho_{f} C_{f}\vec{v}T)=\nabla \cdot(\kappa_{m} \nabla T),
\end{equation}
where $ T $ is  temperature, and $ \rho_{s} $ is the density of solid in porous media, respectively. Here, $ C_{f} $
and $ C_{s} $ are the specific heat of the fluid and solid porous materials, respectively. The effective product of density and specific heat of the whole porous material, governed by the average-volume method~\cite{PJ-Bear2},
\begin{equation}
(\rho C)_{m}=(1-\phi)(\rho_{s} C_{s})+\phi(\rho_{f} C_{f}).
\end{equation}
Similarly, the effective thermal conductivity $ \kappa_{m} $ are also the summation of $ \kappa_{f} $ (for fluids) and $ \kappa_{s} $ (for solids),
\begin{equation}
\kappa_{m}=(1-\phi)\kappa_{s}+\phi\kappa_{f}.
\end{equation}
Note that the Eq.~(\ref{PJ-e2}) is the unsteady convection-diffusion equation whose form-invariance under coordinate transformations are proved~\cite{PJ-GuenneauAIP15,PJ-JRCI2013}. Additionally, the substitution of Eq.~(\ref{PJ-e1}) into Eq.~(\ref{PJ-e2}) leads
\begin{equation}\label{PJ-heat transfer porous2}
(\rho C)_{m}\frac{\partial T}{\partial t}+\rho_{f} C_{f}(\vec{v}\cdot\nabla T)=\nabla \cdot(\kappa_{m} \nabla T).
\end{equation}

Like the steady-state scenarios, we can see all governing equitations satisfy the transformation theory.~\cite{PJ-DaiPRE18}. For both these steady and unsteady situations, the transformation matrix for an isotropic virtual space is as follows:
$ \frac{\textbf{J}\textbf{J}^{\intercal}}{\det\textbf{J}}, $
where $ \textbf{J} $ signifies the Jacobian matrix mapping from the transformed coordinate to its original counterpart. It's also essential to adjust the permeability and heat conductivity as $ \beta'=\dfrac{\textbf{J} \beta \textbf{J}^{\intercal}}{\det\textbf{J}} $ and $ \kappa_{m}'=\dfrac{\textbf{J} \kappa_{m}\textbf{J}^{\intercal}}{\det\textbf{J}} $. Distinctively, for the unsteady scenarios, it becomes necessary to modify both the porosity and the product of density and specific heat. The transformation is
\begin{equation}\label{PJ-e3}
\left\{\begin{array}{ll}\phi'=\dfrac{\phi}{\det\textbf{J}}\\
(\rho_{f} C_{f})'=\rho_{f} C_{f}\\
(\rho_{s} C_{s})'=\dfrac{1-\phi}{\det\textbf{J}-\phi}\rho_{s} C_{s}\end{array}\right.
\end{equation}
Using this approach, we maintain the properties of fluid unchanged, focusing solely on crafting the required solid metamaterial. It's noteworthy that $ \rho_{f} $ is not a constant anymore if we further consider its correlation with temperature variations over time and space. For clarity, we make the assumption that
\begin{equation}
\rho=\rho_{0}[1-\gamma(T-T_{0})],
\end{equation}
where $ \gamma=(\frac{\partial \rho}{\partial T})_{p}/\rho $ represents the density expansion ratio at a constant pressure. Without loss of generality, we give the $ \gamma>0. $ Therefore, we consider the transformed equations:
\begin{equation}
\left\{\begin{array}{ll}\vec{v}'=-\dfrac{\beta'}{\eta}\nabla p\\
\dfrac{\partial (\phi'\rho_{f})}{\partial t}+\nabla\cdot(\rho_{f}\vec{v}')=0\\
(\rho C)'_{m}\dfrac{\partial T}{\partial t}+\rho_{f} C_{f}(\vec{v}'\cdot\nabla T)=\nabla \cdot(\kappa'_{m} \nabla T)
\end{array}\right.,
\end{equation}
where the transformed velocity $ \vec{v}' $ is $\textbf{J}\vec{v}/\det\textbf{J} $ ~\cite{PJ-JRCI2013,PJ-GuenneauAIP15,PJ-DaiPRE18} and $ (\rho C)'_{m}=(1-\phi')(\rho_{s} C_{s})'+\phi'(\rho_{f} C_{f}). $

\section{Potential Applications}

Actually, metamaterials designed by transformation thermotics typically exhibit characteristics that are anisotropic, inhomogeneous, and at times even singular. These traits present significant fabrication challenges. For addressing these complexities in hybrid thermal systems, researchers turn to effective medium theories and multilayered composite structures to achieve the desired outcomes. Regrettably, a fitting theory for managing such hybrid thermal systems has not been developed yet. As a result, there is a pressing need to devise a theory that streamlines the intricate parameters introduced by transformation thermotics.

To address this challenge, we draw inspiration from the concept of neutral inclusion within porous materials. By tailoring two pivotal parameters—thermal conductivity and permeability—we are able to achieve three distinct types of thermal illusions: transparency, concentration, and cloaking. To elaborate, thermal transparency involves creating a core-shell structure to preserve the temperature, velocity, and heat flux distributions of the background undisturbed. Notably, this approach obviates the need for anisotropy, inhomogeneity, and singularity. Similarly, to attain thermal concentration or cloaking, we fashion an anisotropic shell, eliminating the necessity for inhomogeneous and singular parameters. As these three functions—transparency, concentration, and cloaking—maintain the background's temperature, velocity, and heat flux distributions undisturbed, we conveniently term them collectively as thermal illusion.

We assume a steady-state thermal convection-diffusion process in porous materials with incompressible fluids and neglect the viscous dissipation term. Therefore, the governing equation is expressed as
\begin{eqnarray}
\rho_{f}C_{p,f}(\vec{v}\cdot\nabla T)=\nabla\cdot(\overset{\leftrightarrow}\kappa\cdot\nabla T),\label{PJE1}
\end{eqnarray}
where $\rho_{f}$, $C_{p,f}$, and $\vec{v}$ are the density, heat capacity, and the velocity of the fluid at constant pressure, respectively, and $T$ denotes the temperature when the porous material reach equilibrium. Moreover, $\overset{\leftrightarrow}\kappa$, representing the average thermal conductivity tensor of the solid and the fluid material, is defined as $\overset{\leftrightarrow}\kappa = (1 - \phi)\overset{\leftrightarrow}\kappa_{s} + \phi\overset{\leftrightarrow}\kappa_{f}$, where $\phi$ stands for the porosity of the media. $\overset{\leftrightarrow}\kappa_{s}$ and $\overset{\leftrightarrow}\kappa_{f}$ are the thermal conductivity tensors of the solid and the fluid material, respectively. In cases where the fluid exhibits laminar flow at a minimal velocity, the velocity $\vec{v}$ is described by Darcy's law,
\begin{equation}\label{PJE2}
\vec{v}=-\left(\overset{\leftrightarrow}\sigma/\eta\right)\cdot\nabla p,
\end{equation}
where $\overset{\leftrightarrow}\sigma$ and $\eta$ is the permeability tensor and dynamic viscosity, respectively. $p$ stands for pressure. In such case, both Re (Reynolds number) and $\overset{\leftrightarrow}\sigma$ are small enough. The conductive flux $\vec{j}$ is governed by Fourier's law,
\begin{equation}\label{PJE3}
\vec{j}=-\overset{\leftrightarrow}\kappa\cdot\nabla T.
\end{equation}
For clarity, we consider the steady state
\begin{eqnarray}
\nabla\cdot\vec{v}=0,\label{PJE4}\\
\nabla\cdot\vec{j}=0.\label{PJE5}
\end{eqnarray}
We also consider one type of fluid with constant dynamic viscosity $\eta$. Then, Eqs.~(\ref{PJE4}) and (\ref{PJE5}) are expressed as
\begin{eqnarray}
\nabla\cdot(-\overset{\leftrightarrow}\sigma\cdot\nabla p) = 0, \label{PJE8}\\
\nabla\cdot(-\overset{\leftrightarrow}\kappa\cdot\nabla T) = 0. \label{PJE9}
\end{eqnarray}
Finally, Eqs.~(\ref{PJE8}) and (\ref{PJE9}) share a comparable mathematical structure. Therefore, the effective medium theory is capable of addressing both thermal conductivity and permeability. Using $\tau$, we can harmonize the representation of $\kappa$ and $\sigma$.

\begin{figure}[!ht]
\includegraphics[width=\linewidth]{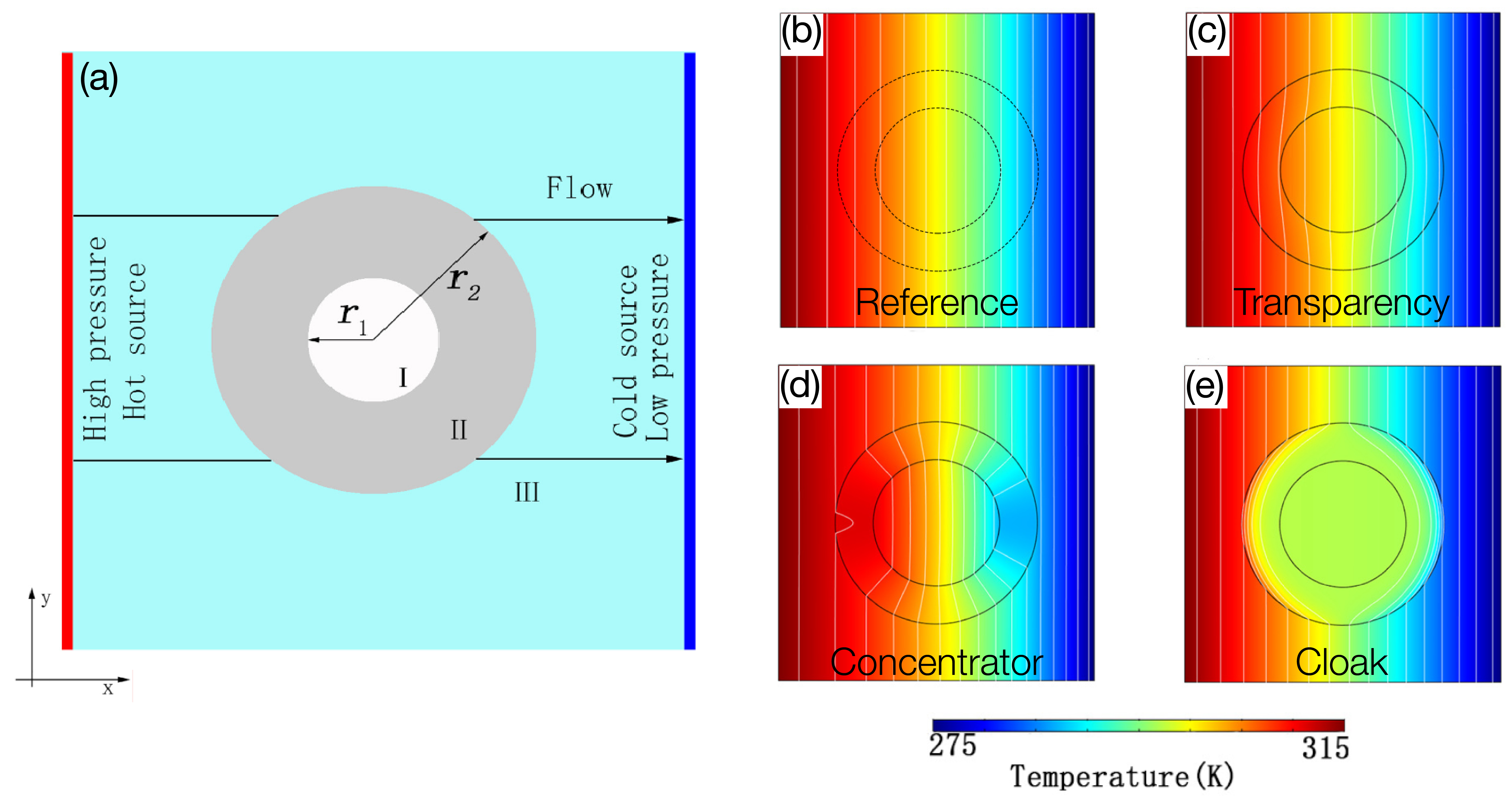}
\caption{Thermal illusion in porous materials. The scale of the system is $10^{-5}$~m. {\bf a} Schematic. The background velocity is along the $x$ direction, as described by the black flow lines. Region I $(r<r_{1})$ is composed of isotropic porous media, region II $(r_{1}<r<r_{2})$ is composed of isotropic media for thermal transparency, and anisotropic media for thermal concentrator or cloak, and region III $(r>r_{2})$ is composed of isotropic background porous media. For thermal illusion in porous media, the black lines in region III with the core-shell structure should be undistorted. {\bf b} Temperature profile of a pure background (reference). {\bf c} Temperature profile of a thermal transparency. {\bf d} Temperature profile of a thermal concentrator. {\bf e} Temperature profile of a thermal cloak. White lines are the isotherms. Adapted from Ref.~\cite{PJ-YangESEE19}}
\label{PJ-1}
\end{figure}

We aspire to eliminate the need for anisotropic, inhomogeneous, and singular parameters. To this end, we turn to the concept of neutral inclusion. This idea provides a methodology to determine the effective thermal conductivity of a core-shell configuration. Subsequently, it becomes essential to compute the effective permeability for the same structure. As depicted in Fig.~\ref{PJ-1}, we designate the core to be isotropic with parameter $\tau_{1}$, the metashell to be anisotropic with parameter $\overset{\leftrightarrow}\tau_{2} = {\rm diag}(\tau_{rr},\tau_{\theta\theta})$ (the shell becomes isotropic when $\tau_{rr} = \tau_{\theta\theta}$ ), and the background to be isotropic with parameter $\tau_{3}$. Consequently, the core-shell structure's effective parameter $\tau_{e}$ can be deduced as follows:
\begin{eqnarray}
\tau_{e}=c\tau_{rr}\frac{\tau_{1}+c\tau_{rr}+(\tau_{1}-c\tau_{rr})f^{c}}{\tau_{1}+c\tau_{rr}-(\tau_{1}-c\tau_{rr})f^{c}},\label{PJ-requirement}
\end{eqnarray}
where $c=\sqrt{\tau_{\theta\theta}/\tau_{rr}}$ denotes the anisotropy of the shell, and $f=(r_{1}/ r_{2})^{2}$ represents the core fraction. To maintain the heat flux and velocity distributions in the background (region III) as though the core-shell structure is absent at the center, we define $\tau_{e}=\tau_{3}$.

Next, we perform finite-element simulations to validate the theory. As shown in schematic of Fig.~\ref{PJ-1}, we take the pressure source as $\Delta p = 400$~Pa and the heat source $\Delta T = 40$~K. The liquid in porous material is set as water with $\rho_{f} = 10^{3}$~kg/m$^{3}$, $C_{p,f} = 4.2\times10^{3}$~J$\cdot$kg$^{-1}$K$^{-1}$, the dynamic viscosity $\eta = 10^{-3}$~Pa$\cdot$s, and $\kappa_{f} = 0.6$~Wm$^{-1}$K$^{-1}$. The porosity is $\phi=0.9$. The size parameters are $r_{1} = 2\times10^{-5}$~m and $r_{2} = 3.2\times10^{-5}$~m. The average thermal conductivity tensors are set to be $\kappa_{1}=6$~Wm$^{-1}$K$^{-1}$, $\overset{\leftrightarrow}\kappa_{2} = {\rm diag}(4,4)$~Wm$^{-1}$K$^{-1}$, and $\kappa_{3} = \kappa_{e}$ given by Eq.~(\ref{PJ-requirement}). The thermal conductivity tensor of the solid are calculated as $\overset{\leftrightarrow}\kappa_{s} = (\overset{\leftrightarrow}\kappa - \phi\kappa_{f})/(1 - \phi)$. The permeability tensors are set to be $\sigma_1 = 5\times10^{-12}$~m$^{2}$, $\overset{\leftrightarrow}\sigma_{2} = {\rm diag}(2,2)\times10^{-12}$~m$^{2}$ (the magnitude $10^{-12}$ is common in nature), and $\sigma_{3} = \sigma_{e}$ given by Eq.~(\ref{PJ-requirement}). In all these cases, we calculate Reynolds numbers ${\rm Re} = r_{2}\rho_{f}v/\eta<1$ (the maximum value is 0.64) and $\sigma\ll r_{2}^{2}$, ensuring the applicability of Darcy's law. The simulation results are shown in Figs.~\ref{PJ-1}. Figure~\ref{PJ-1}b displays a simulation of a pure background without the core-shell structure, serving as a reference. Based on the above core-shell parameters, we then produce a thermal transparency pattern; see Fig.~\ref{PJ-1}c. For realizing a thermal cloak (or concentrator) pattern, we set $\tau_{rr} \ll \tau_{\theta\theta}$ (or $\tau_{rr} > \tau_{\theta\theta}$). Temperature profiles are shown in Fig.~\ref{PJ-1}d and Fig.~\ref{PJ-1}e. Finally, different thermal patterns are created including thermal transparency, thermal cloak and thermal concentrator. The performance of thermal illusion are not affected as long as the permeability and thermal conductivity satisfy Eq.~(\ref{PJ-requirement}).

\section{Laboratory Experiment of Steady-state Transformation Thermo-hydrodynamics}

\begin{figure}[t]
\includegraphics[width=1.0\linewidth]{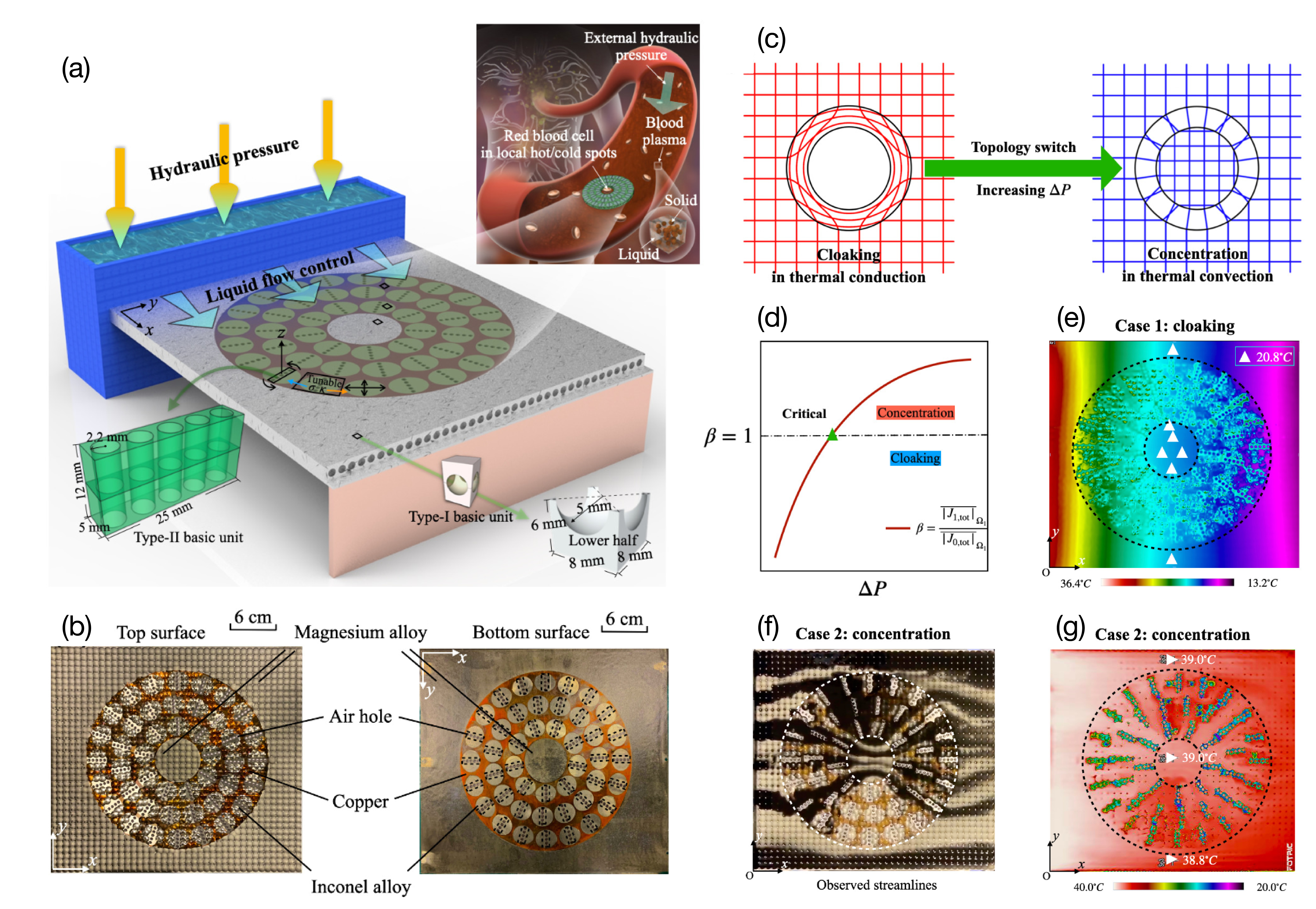}
\caption{Liquid-solid hybrid thermal metamaterial. {\bf a} Illustration of the metadevice based on the liquid-solid hybrid thermal metamaterial. {\bf b} Photos of the top and the bottom of the sample. Scale bar is 6~cm. {\bf c} The switch between thermal cloaking and thermal concentration corresponds to a topological switch in virtual space. {\bf d} The heat flux amplification factor $\beta$ can be tuned continuously by the external hydraulic pressure. Meanwhile, the function of the metadevice is switched. {\bf e} Measured temperature profile of the thermal metadevice at $\Delta P$ = 0. White triangles denote the positions with the temperature of 20.8~$^{\circ}C$. {\bf f} Observed streamlines of the thermal metadevice at $\Delta P$ $\ne$ 0. {\bf g} Measured temperature profile of the thermal metadevice at $\Delta P$ $\ne$ 0. Horizontal white triangles denote the positions with the temperature (from up to down) of 39.0~$^{\circ}C$, 39.0~$^{\circ}C$, and 38.8~$^{\circ}C$, respectively. Adapted from Ref.~\cite{PJ-JinPNAS23}}
\label{PJ-F1}
\end{figure}

In crafting the hybrid thermal metamaterial, we employ fine-designed porous structures, enabling both thermal convection and conduction to coexist within the same space, as illustrated in Fig.~\ref{PJ-F1}a. This design process is bifurcated into two stages. Initially, by sculpting the basic unit, we engineer a porous substance that allows localized, independent modulation of both thermal conduction and convection attributes. Subsequently, leveraging the advanced principles of transformation thermotics, we shape the spatial characteristics of these thermal properties to realize the intended functionalities of the thermal metadevice.

The local manipulation of both conductive and convective thermal properties is achieved through the design of basic units. We have two types of units. The type-I unit comprises a cuboid featuring a hemispherical region filled with water (see lower-right inset of Fig.~\ref{PJ-F1}a). In contrast, the type-II unit is a cuboid possessing cylindrical five air holes. The effective thermal conductivity of each unit is given by $\bm{\kappa}=\left(1-\phi_l-\phi_a\right)\bm{\kappa}_{\rm s}+\phi_l\bm{\kappa}_{\rm l} + \phi_a\bm{\kappa}_{\rm a}$ where $\bm{\kappa}_{\rm s}$, $\bm{\kappa}_{\rm l}$, and $\bm{\kappa}_{\rm a}$ are the thermal conductivity of the solid, liquid, and air, respectively. $\phi_l$ and $\phi_a$ represent the filling fraction of the liquid and air region, respectively. Within each unit, the thermal conductivity can be adjusted based on the selected solid material and the filling fractions $\phi_l$ and $\phi_a$. Concurrently, the permeability $\bm{\sigma}$ can be modulated based on the geometry of the liquid or air region. For instance, in the type-II units air holes are employed to deflect the liquid flow. The orientation of such units can be strategically adjusted to modify the permeability $\bm{\sigma}$.

We craft a metashell design wherein the thermal conductivity $\bm{\kappa}'$ distribution is tailored for thermal cloaking, guided by our choice of the transformation $\bm{\Xi}$. This transformation correlates to a virtual space with a hole at the center. The hole in the virtual space is exactly the origin of the thermal cloaking effect: The heat flows cannot touch any object in the hole in the virtual space, while in real space, an object in the core region remains unaffected by the heat flows. Conversely, the liquid permeability distribution $\bm{\sigma}'$ arises from the transformation of the thermal convection $\bm{\Lambda}$, crafted for thermal concentration. This transformation maps to a virtual space with no hole. From the geometric point of view, the virtual space with a hole is topologically distinct from the virtual spaces with no hole. Consequently, with increasing hydraulic pressure difference $\Delta P$ = $P_{\rm h}$ - $P_{\rm l}$ ($P_{\rm h}$ and $P_{\rm l}$ are the hydraulic pressure at the hot and cold sides of the metadevice, respectively), thermal convection becomes dominant and the device function switches from thermal cloaking to thermal concentration. Meanwhile, the virtual space undergoes topology switch (see Fig.~\ref{PJ-F1}c). In particular, the nontrivial topology within the virtual space for thermal cloaking indicates that there are some properties robust to external conditions. These properties are the heat current in the core region. In the thermal cloaking regime, such a heat current is irrelevant with external temperature distributions. In contrast, for thermal concentration, the heat current in the core region is highly sensitive to external temperature regions. The switch between these two functions reflect the topology change in the virtual space. To provide a quantitative assessment of our metadevice's function, we introduce the heat flux amplification factor $\beta$. This is determined by the averaged amplitude of the total heat flux in the core region ($\Omega_1$) over the same quantity when the system is changed to the background (henceforth denoted as ``the reference'').

We proceed to showcase the transition between thermal cloaking and thermal concentration through regulated hydrodynamics. Under boundary condition I, where $\Delta P$ = 0, we evaluate the temperature profile within the metadevice. As depicted in Fig.~\ref{PJ-F1}e, the temperature distribution captured by the infrared camera displays a perfect pattern of thermal cloaking. Notably, the core region has a consistent temperature distribution around $20.8^\circ C$. This suggests no conductive heat flow in the core region. Additionally, the temperature profile in the background region remains mostly undisturbed. Therefore, it can be interpreted as $\beta < 1$, signifying that the metadevice is operating in the cloaking mode.

Under boundary condition II, we target for the realization of the thermal concentration. We ensure that thermal convection takes precedence in these scenarios. We experimentally showcase the performance of thermal concentration, examining it through the lens of fluid dynamics and temperature profiling. To visualize the fluid flow, we perforate six holes beneath a colorant container, a concoction of alkanes and toner. Once the system stabilizes into a nonequilibrium steady-state, the colorant is methodically dripped from the metadevice's left boundary through these holes, ensuring simultaneous and equidistant distribution. Fig.~\ref{PJ-F1}f showcases the six streamlines. The central four streamlines converge into the core area, and all the streamlines outside the region $\Omega_2$ are only marginally distorted. This pattern underscores that the core zone experiences a more substantial flow (or a heightened fluid velocity) compared to the backdrop. It is worth noting that within the $\Omega_2$ region, colorant distribution differs between the upper and lower sections. This discrepancy arises from the asymmetrical layout of the type-II basic units, further compounded by the unequal positioning of the six holes in relation to these sections. Yet, the fluid flow's concentration remains distinctly visible in Fig.~\ref{PJ-F1}g. Subsequently, we gauge the temperature profile in the metadevice under identical conditions. The measured temperature profile (see Fig.~\ref{PJ-F1}g) exhibits several features. First, the overall temperature of the metadevice is higher than in the cloaking case. Moreover, the temperature gradient is pushed to the right side of the metadevice. There are visible correlations between the temperature profile and liquid flow profile, indicating that the thermal transport is now dominated by the convective heat flow carried by the water. The convective heat flow in the core region is larger than that in the background region because of $v_{\Omega_1} > v_{\Omega_3}$ with $T_{\Omega_1}$ $\approx$ $T_{\Omega_3}$ (see the white triangles in Fig.~\ref{PJ-F1}g). In this phase, $\beta > 1$. Consequently, the metadevice is transitioned into thermal concentration by increasing the external hydraulic pressure.

\section{Discussion and Conclusion}

In conclusion, we delve into the transformation thermo-hydrodynamics theory and the corresponding experimental methodologies that probe heat transfer in porous materials. These specially engineered porous materials exhibit remarkable adaptability in heat management, all while ensuring the temperature field in the backdrop remains undisturbed — a feat beyond the capabilities of traditional thermal metamaterials. Characterized by their unique liquid-solid amalgamation, these hybrid thermal metamaterials hold immense promise for a plethora of applications. These include thermal illusions and camouflage, enhanced cooling and heat regulation in electronic gadgets, sustainable infrastructure, and sophisticated heat modulation in intelligent materials and machinery. Delving deeper into these hybrid metamaterials might just unveil groundbreaking insights into the intricacies of complex systems.~\citep{PJ-ZhouPA2009,PJ-ZhangCPL2009,PJ-ZhaoJAP2009,PJ-TanJPCB2009,PJ-TianJAP2009,PJ-GuCTP2009,PJ-WangOL2009,PJ-GaoJAP2009,PJ-TanJPCB2009-1,PJ-LiuEPJAP2009,PJ-HanPLA2009,PJ-JianJPCC2009,PJ-FanJPCC2009,PJ-WangPNAS09,PJ-WuEPJAP09,PJ-HuangSSC00,PJ-HuangCTP01-2,PJ-HuangCTP01-1,PJ-PanPB01,PJ-HuangPRE01,PJ-HuangJPCM02,PJ-HuangPRE02,PJ-HuangCTP02,PJ-HuangPLA02,PJ-HuangCTP03,PJ-HuangPRE03-2,PJ-GaoPRE03,PJ-HuangJAP03,PJ-HuangPRE03-1,PJ-GaoEPJB03,PJ-DongJAP04,PJ-KoJPCM04,PJ-GaoPRB04,PJ-HuangPRE04g,PJ-LiuPLA04,PJ-DongJAP04-1,PJ-HuangCPL04,PJ-HuangPRE04f,PJ-KoEPJE04,PJ-HuangAPL04,PJ-HuangCP04,PJ-HuangPRE04e,PJ-HuangEL04,PJ-HuangJPCB04,PJ-HuangJCP04,PJ-HuangPRE04d,PJ-HuangPRE04c,PJ-HuangJPCM04,PJ-HuangPLA04}, including nonlinear systems~\citep{PJ-ZhangCPB10,PJ-LiuCTP10,PJ-FanCTP10,PJ-GaoPRL10,PJ-WangOL10,PJ-BaoJPCM10,PJ-GaoPP10,PJ-TanSM10,PJ-MengJPCB11,PJ-SuFOP11,PJ-LiuIJMPB11,PJ-ZhaoPNAS11,PJ-FanJPDAP11,PJ-WangJPCB11,PJ-ZhaoFOP12,PJ-SongARCS12,PJ-Li2012sm,PJ-Wang2012cpb,PJ-li2012epl,PJ-huang2013cp,PJ-wei2013plos,PJ-liang2013pre,PJ-Fan2013fop,PJ-Meng2013mp,PJ-Wang2013ACM,PJ-Liu2013ASP,PJ-wei2013AnAM,PJ-chen2013prl,PJ-Fan2013cpb,PJ-Liang2013fopw,PJ-Meng2013pre,PJ-Li2013EPJP,PJ-QiuEPJB13,PJ-QiuPLoS13,PJ-QiuFP14,PJ-QiuPLA14,PJ-QiuPLoS14,PJ-QiuCPB14,PJ-QiuJSM14,PJ-QiuCTP14,PJ-QiuPLoS14-1,PJ-QiuTEPJ-AP14,PJ-QiuCTP15,PJ-QiuSpringer15,PJ-QiuJP.Phys.Chem.B15,PJ-QiuPR15,PJ-QiuEur.Phys.J.Appl.Phys.15,PJ-QiuPLA15,PJ-Tan19,PJ-Tan18,PJ-Tan17,PJ-Tan12,PJ-Tan10,PJ-Tan9,PJ-Tan8}, soft matter systems~\citep{PJ-HuangPRE04b,PJ-HuangPRE04a,PJ-HuangJAP05,PJ-XiaoPRB05,PJ-HuangAPL05b,PJ-HuangOL05,PJ-HuangJPCB05,PJ-LiuCTP05,PJ-HuangPLA05,PJ-HuangPRE05b,PJ-HuangPRE05a,PJ-HuangJOSAB05,PJ-HuangAPL05a,PJ-DongEPJB05,PJ-HuangJMMM05,PJ-HuangJAP06,PJ-TianPRE06,PJ-WangCPL06,PJ-ShenCPL06,PJ-WangJPCB06,PJ-CaoJPCB06,PJ-HuangPR06a,PJ-XuPLA06,PJ-TianCPL06,PJ-FanAPL06,PJ-XueCPL06,PJ-FanJPCB06,PJ-TianPRE07,PJ-YangJAP07,PJ-TianEPL07,PJ-FangCPL07,PJ-ChenJPA07,PJ-GaoJPCC07,PJ-WangAPL07,PJ-HuangNY07,PJ-ZhuJAP08,PJ-YePA081,PJ-HuangJRCC08,PJ-XuJMR08,PJ-JianJRCB08,PJ-ZhangAPL08,PJ-GaoAPL2008,PJ-FanJAP2008,PJ-XiaoJPCB2008,PJ-2008,PJ-WangOL2008,PJ-Tan3,PJ-Tan2}, and statistical physics~\cite{PJ-Tan14,PJ-Tan13,PJ-Tan7,PJ-Tan6,PJ-Tan5}.

Let's delve into the topological transition inherent to the hybrid metamaterial. From a physics standpoint, the transformation theory serves as a conduit, bridging geometric transformations with their parametric counterparts. As a result, the topological attributes of a geometric transformation find their mirrored representation in material parameters. This allows us to explore topology via (effective) thermal conductivity. For instance, when considering a core-shell structure designed for thermal cloaking, its effective thermal conductivity can be perceived as a topological invariant — it remains unaltered by the thermal conductivity of the core. This implies that a geometric transformation with a distinct topological nature can lead to singular (either zero or infinite) thermal conductivity, effectively segmenting a region from its surroundings. Hence, transformation thermotics becomes instrumental in drawing connections between geometric and parametric topologies. Broadly speaking, situations with zero or infinite thermal conductivity bear topological significance. This is analogous to the unique nature of zero-index metamaterials: they facilitate seamless transport without reflections in real space and correspond to materials showcasing Dirac dispersions in wave-vector space.

In wrapping up, we foresee a transformative application of our metadevice in managing localized temperature variations within biological cells or tissues~\cite{PJ-TY}, all while ensuring the broader environment of the human body remains unperturbed. The extracellular fluid within organisms, a composite of liquid and solid-like elements (as illustrated in Fig.~\ref{PJ-F1}a), can be conceptualized as a porous medium. By designing porous metamaterials tailored to a specific scale, we can effectively model this environment. In a consistent external setting, discerning the direction of localized heat flow is feasible. Given our structure's axisymmetric design, as guided by the transformation theory, there's no external deviation. This implies that our model can be oriented along the direction of this localized heat flow. The adiabatic boundaries positioned at the top and bottom mirror open boundaries since there's a null heat flow vertically. This makes our model an apt representation of real conditions. By manipulating external hydraulic pressures, our uniquely designed metamaterials can regulate localized temperature variations in living cells or tissues. These hot/cold spots typically correspond to temporary temperature deviations. Harnessing hydraulic pressure control, these deviations can be mitigated swiftly. For such localized temperature variances, elevating fluid velocity and heat flow can expedite the shift towards a more stable temperature environment. Moreover, within a biological context, this supplemental flux aids in balancing chemical concentrations, such as ATPs and CO$_2$, further bolstering the restoration of biological functions. Furthermore, embedding these micro-scale metamaterials within the human body for therapeutic purposes is a tangible possibility. Our design's capability to operate without disrupting the background thermal or fluidic environment is of paramount significance, especially given its implications for human health.

\end{document}